\documentclass[sigconf, nonacm]{acmart}
\usepackage{multirow}
\usepackage{algorithm}
\usepackage[algo2e,ruled,vlined]{algorithm2e}
\usepackage{graphicx}
\usepackage{amsmath}
\usepackage{array}
\usepackage{graphicx}
\usepackage{booktabs}
\usepackage{multirow}
\usepackage{chngpage}
\usepackage{bbm}
\usepackage{enumitem}
\usepackage{subcaption}
\usepackage{amsmath}
\usepackage{xcolor}
\definecolor{ForestGreen}{RGB}{34,160,34}
\usepackage{tikz}

\usepackage{booktabs}
\usepackage{multirow}
\usepackage{colortbl}
\usepackage{xcolor}
\usepackage{graphicx}
\usepackage{subcaption}
\usepackage{tabularx}

\definecolor{grayrow}{gray}{0.9}
\newcommand{\mypara}[1]{\vspace{2.5pt}\noindent\textbf{#1.}}

\newcommand{\cutsectionup}{\vspace*{-0.00in}}

\newcommand{\cutcaptionup}{\vspace*{-0.08in}}

\newcommand{\model}[1]{\textsf{\small{ERA}}}

\renewcommand{\eqref}[1]{\ref{#1}}

\AtBeginDocument{%
  \providecommand\BibTeX{{%
    \normalfont B\kern-0.5em{\scshape i\kern-0.25em b}\kern-0.8em\TeX}}}

\copyrightyear{2026}
\acmYear{2026}
\setcopyright{acmlicensed}
\acmConference[KDD '26] {Proceedings of the ACM SIGKDD Conference on Knowledge Discovery and Data Mining}{XX XX-- XX XX, 2026}{XXX, XXX.}
\acmBooktitle{Proceedings of the ACM SIGKDD Conference on Knowledge Discovery and Data Mining (KDD 26), August 9-- August 13, 2026, Jeju, Korea}
\acmPrice{15.00}
\acmDOI{xx.xxxx/xxxxxxx.xxxxxxx}
\acmISBN{xxx-x-xxxx-xxxx-x/xx/xx}

\begin{document}
\title{ERA: Evidence-based Reliability Alignment for Honest Retrieval-Augmented Generation}
\settopmatter{authorsperrow=4, printacmref=true}

\makeatletter
\renewcommand*{\@fnsymbol}[1]{\ensuremath{\ifcase#1\or \dagger\or \ddagger\or \mathsection\or \mathparagraph\or \|\or **\or \dagger\dagger \or \ddagger\ddagger \else\@ctrerr\fi}}
\makeatother

\author{Sunguk Shin}
\orcid{0000-0002-3424-405X}
\affiliation{%
  \institution{Korea University}
  \city{Seoul}
  \country{South Korea}
}
\affiliation{%
  \institution{MPI-SP}
  \city{Bochum}
  \country{Germany}
}
\email{ssw1419@korea.ac.kr}

\author{Meeyoung Cha}
\orcid{0000-0003-4085-9648}
\affiliation{%
  \institution{MPI-SP}
  \city{Bochum}
  \country{Germany}
}
\affiliation{%
  \institution{KAIST}
  \city{Daejeon}
  \country{South Korea}
}
\email{mia.cha@mpi-sp.org}

\author{Byung-Jun Lee}
\authornote{Co-corresponding authors}
\orcid{0000-0002-0684-607X}
\affiliation{%
 \institution{Korea University}
 \city{Seoul}
 \country{South Korea}
}
\affiliation{%
 \institution{Gauss Labs Inc.}
 \city{Seoul}
 \country{South Korea}
}
\email{byungjunlee@korea.ac.kr}

\author{Sungwon Park}
\authornotemark[1]
\orcid{0000-0002-6369-8130}
\affiliation{%
  \institution{KAIST}
  \city{Daejeon}
  \country{South Korea}
}
\affiliation{%
  \institution{MPI-SP}
  \city{Bochum}
  \country{Germany}
}
\email{psw0416@kaist.ac.kr}

\begin{abstract}
Retrieval-Augmented Generation (RAG) grounds language models in factual evidence but introduces critical challenges regarding knowledge conflicts between internalized parameters and retrieved information. However, existing reliability methods, typically relying on scalar confidence, fail to explicitly distinguish between epistemic uncertainty and inherent data ambiguity in such hybrid scenarios. In this paper, we propose a new framework called \model{} (Evidence-based Reliability Alignment) to enhance abstention behavior in RAG systems by shifting confidence estimation from scalar probabilities to explicit evidence distributions. Our method consists of two main components: (1) Contextual Evidence Quantification, which models internal and external knowledge as independent belief masses via the Dirichlet distribution, and (2) Quantifying Knowledge Conflict, which leverages Dempster-Shafer Theory (DST) to rigorously measure the geometric discordance between information sources. These components are used to disentangle epistemic uncertainty from aleatoric uncertainty and modulate the optimization objective based on detected conflicts. Experiments on standard benchmarks and a curated generalization dataset demonstrate that our approach significantly outperforms baselines, optimizing the trade-off between answer coverage and abstention with superior calibration.

\end{abstract}

\begin{CCSXML}
<ccs2012>
   <concept>
       <concept_id>10010147.10010178.10010179.10010182</concept_id>
       <concept_desc>Computing methodologies~Natural language generation</concept_desc>
       <concept_significance>500</concept_significance>
   </concept>
   <concept>
       <concept_id>10002951.10003317.10003347.10003348</concept_id>
       <concept_desc>Information systems~Question answering</concept_desc>
       <concept_significance>500</concept_significance>
   </concept>
   <concept>
       <concept_id>10010147.10010257.10010293.10010294</concept_id>
       <concept_desc>Computing methodologies~Neural networks</concept_desc>
       <concept_significance>300</concept_significance>
   </concept>
   <concept>
       <concept_id>10010147.10010257.10010321.10010336</concept_id>
       <concept_desc>Computing methodologies~Feature selection</concept_desc>
       <concept_significance>100</concept_significance>
       </concept>
   <concept>
       <concept_id>10010147.10010257.10010258.10010259</concept_id>
       <concept_desc>Computing methodologies~Supervised learning</concept_desc>
       <concept_significance>300</concept_significance>
   </concept>
</ccs2012>
\end{CCSXML}

\ccsdesc[500]{Computing methodologies~Natural language generation}
\ccsdesc[500]{Information systems~Question answering}

\keywords{}

\maketitle

\cutsectionup
\section{Introduction}
Recent advances in Large Language Models (LLMs), trained on massive corpora, have profoundly transformed the landscape of Natural Language Processing (NLP). Despite their remarkable capabilities, these models remain prone to hallucinations, stemming from their reliance on static training data that may become outdated and their tendency to generate factually incorrect or misleading content~\cite{kalai2025language,zhang2024language}. Retrieval-Augmented Generation (RAG) has emerged as a promising solution by incorporating external knowledge sources into the generation process, thereby grounding model outputs in factual evidence and improving overall reliability~\cite{lewis2020retrieval}.

However, despite its advantages, RAG introduces a new class of challenges arising from its hybrid nature. In particular, conflicts may emerge between a model’s internalized parametric knowledge and the retrieved external information~\cite{longpre2021entity,mallen2023not}. Such discrepancies can create ambiguity during generation, making it difficult for the model to reconcile conflicting evidence. As a result, the model may exhibit inconsistent behavior, degraded response quality, or erroneous reasoning when it fails to appropriately prioritize the correct information source~\citep{shin2025enhancing}. Addressing this challenge is critical for ensuring the trustworthiness of retrieval-augmented systems in real-world applications.

To mitigate these issues, prior work has explored mechanisms for regulating a model’s reliance on retrieved information, including dynamic gating strategies. In parallel, early research investigated unsupervised approaches for assessing response reliability through statistical proxies. For example, Guerreiro et al. leveraged token-level output probabilities—such as average or minimum logit values—to estimate generation confidence~\cite{guerreiro2023looking}, while Wang et al. proposed Self-Consistency, which samples multiple reasoning paths and selects the most reliable answer via majority voting~\cite{wangself}. In contrast, another line of work adopts a supervised paradigm. For instance, Sun et al. employed Direct Preference Optimization (DPO) to train models to abstain from answering when retrieved context is insufficient, using curated abstention datasets to align model behavior with reliability objectives~\cite{sun2025divide}.

Despite these advances, existing approaches remain vulnerable to overfitting—either to the pre-training distribution in unsupervised methods or to specific abstention patterns in supervised ones. As a result, unsupervised methods often struggle to distinguish confident hallucinations from correct predictions, while supervised approaches may memorize refusal behaviors, limiting their generalization to unseen scenarios. These shortcomings stem from two fundamental challenges. First, conventional probabilistic frameworks rely on scalar confidence estimates that fail to distinguish epistemic uncertainty (arising from lack of knowledge) from aleatoric uncertainty (stemming from inherent data ambiguity), often leading to overconfident predictions~\cite{guo2017calibration,kuhnsemantic,park2025enhancing}. Second, the aggregation of internal model knowledge and external retrieved contexts into a single output distribution obscures conflicts between these information sources, hindering the reliable identification of hallucinations~\cite{mallen2023not,jin2025disentangling}. This highlights the critical need for a robust quantification metric capable of explicitly disentangling belief from uncertainty while simultaneously measuring the conflict between internal parametric knowledge and external retrieved evidence.

To address these limitations, we propose \model{} (Evidence-based Reliability Alignment), a principled framework designed to enhance abstention behavior in RAG systems. Inspired by evidential deep learning~\cite{sensoy2018evidential}, \model{} shifts confidence estimation from scalar probabilities to explicit modeling of evidence distributions. By leveraging the Dirichlet distribution to parameterize evidence density, our approach disentangles epistemic and aleatoric uncertainty in a principled manner. Furthermore, \model{} independently models belief masses derived from parametric knowledge and retrieved evidence. Using Dempster–Shafer Theory (DST)~\cite{jsang2018subjective}, it quantitatively measures belief conflict—the disagreement between internal priors and external information—thereby providing a rigorous geometric foundation for reliable and interpretable abstention.

Experiments demonstrate that \model{} optimizes the trade-off between answer coverage and abstention, maintaining high accuracy on answerable queries while selectively withholding responses in uncertain or conflicting scenarios. In comparisons with several recent baselines on standard RAG benchmarks, our results show that the model significantly improves response reliability without compromising its ability to answer supported queries. Furthermore, this robustness is validated through evaluations on a curated dataset of Wikipedia events, specifically designed to assess generalizability. Finally, extensive empirical analysis confirms that \model{} produces well-calibrated confidence estimates, ensuring that its predicted uncertainty accurately reflects the true likelihood of correctness.

\vspace{1mm}
The major contributions and findings of our work include: 
\vspace{1mm}

\begin{itemize} [nosep,leftmargin=1em,labelwidth=*,align=left]

\item Proposing a novel reliability alignment framework, \model{}, that fundamentally shifts confidence quantification from scalar probabilities to evidence distributions via the Dirichlet distribution, explicitly disentangling epistemic uncertainty from aleatoric uncertainty. \looseness=-1

\item Introducing a rigorous quantification of `belief conflict' by leveraging Dempster-Shafer Theory (DST) to measure the geometric discordance between the model's internal parametric knowledge and external retrieved evidence, which are modeled as independent belief masses.

\item Constructing a curated benchmark dataset derived from Wikipedia events specifically designed to evaluate generalization, verifying that \model{} robustly identifies unsupported queries across diverse domains without overfitting to specific refusal patterns.

\item Demonstrating through extensive experiments that \model{} produces well-calibrated confidence estimates and optimizes the trade-off between answer coverage and abstention, significantly outperforming existing baselines in enabling honest RAG. \looseness=-1

\end{itemize}

\vspace{1mm}
\noindent 
The training details and  datasets are publicly available at  \url{https://anonymous.4open.science/r/ERA-318B}.\looseness=-1
\cutsectionup
\section{Related work}
\subsection{Uncertainty Quantification and Evidential Deep Learning}
In deep learning classifiers, softmax probabilities are widely used as a proxy for confidence; however, they are often miscalibrated and can be overconfident~\citep{guo2017calibration}. To mitigate this miscalibration, various approaches have been proposed, ranging from approximate Bayesian inference methods~\citep{gal2016dropout} to Deep Ensembles~\citep{lakshminarayanan2017simple,park2021improving}. While these methods often increase training cost due to sampling or multiple models, this added computational overhead can be undesirable for real-time or large-scale applications. To address this limitation, Evidential Deep Learning (EDL) predicts class-wise evidence and uses it to parameterize a Dirichlet distribution over class probabilities~\citep{sensoy2018evidential}, enabling the model to compute both expected predictive probabilities and a corresponding measure of predictive uncertainty. This evidential approach has been further advanced by methods like Posterior Networks, which leverage latent-space normalizing-flow density estimation to improve the quality of evidence quantification~\citep{charpentier2020posterior}.

Rather than treating class probabilities as a point estimate, EDL represents predictions as a Dirichlet distribution over class probabilities, thereby modeling uncertainty at the distribution level~\citep{sensoy2018evidential}. Concretely, the network outputs non-negative class-wise evidence $e_k \geq 0$, which is mapped to Dirichlet concentration parameters via $\alpha_k = e_k + 1$. This formulation is closely related to Subjective Logic, where an opinion is decomposed into belief masses $\{b_k\}$ and an uncertainty mass $u$, satisfying the conservation constraint $u + \sum_k{b_k} = 1$~\citep{jsang2018subjective}. Recent studies in Trusted Multi-View Classification demonstrate that fusing such evidential opinions from disparate sources significantly enhances reliability~\citep{han2022trusted}. By separating supported belief from residual uncertainty, the evidential parameterization provides a structured representation of "knowledge" versus "ignorance".

To make the ignorance mass meaningful, EDL typically couples a data-fitting term with a regularization term that discourages unwarranted evidence, often implemented via a KL-divergence penalty toward a high-uncertainty reference Dirichlet distribution~\citep{sensoy2018evidential}. This regularization encourages flatter predictive distributions when evidence is weak or ambiguous, reducing unsupported overconfidence and making evidential outputs more suitable for reliability-aware decision rules~\citep{sensoy2021misclassification}.

\cutsectionup
\subsection{Honest RAG and Knowledge Conflicts}
Hallucinations persist even in advanced language models and are not eliminated simply by augmenting generation with external tools like search or retrieval~\citep{kalai2025language,lewis2020retrieval,asai2024self}. Recent analyses argue that hallucinations stem from statistical pressures during pretraining~\citep{kalai2025language}. These tendencies persist post-training because standard optimization rewards guessing over acknowledging uncertainty. Due to the tendency of LLMs to hallucinate~\citep{kadavath2022language,zhang2024language}, the problem is merely mitigated when retrieval is accurate, but persists whenever the retrieved context is imperfect or absent. These observations motivate \emph{honest RAG}, a paradigm focused on determining when retrieval-augmented systems should answer versus abstain~\cite{lin2022truthfulqa}, and how to prevent responses that are unsupported by evidence~\citep{chen2024controlling,wu2024synchronous}.

Within the scope of honest RAG, initial efforts have focused on measuring faithfulness violations. For instance, RAGTruth provides a corpus annotating cases where outputs conflict with or are unsupported by retrieved evidence~\citep{niu2024ragtruth,maynez2020faithfulness}. This perspective shifts the diagnosis of RAG failures from "missing retrieval" to "unused or misused retrieval"~\citep{jiang2023active}. Complementing this measurement-focused view, inference-time decoding interventions suggest adjusting decoding to reduce hallucinations~\citep{chuang2023dola}.

A parallel thread frames honest RAG as a risk-control or alignment problem. RC-RAG advocates for controlling the risk of answering under low confidence, analyzing factors like retrieval quality to motivate selective abstention~\citep{chen2024controlling}. Similarly, Divide-Then-Align (DTA) formalizes the problem by defining boundaries between parametric knowledge and retrieval-provided knowledge~\citep{sun2025divide}. By partitioning data into quadrants (e.g., Known/Unknown Parametric vs. Known/Unknown Context), DTA aligns models to answer when supported and abstain when necessary.

However, a critical gap remains. Existing approaches~\citep{chen2024controlling, sun2025divide} often treat "uncertainty" and "conflict" as outcomes to be detected or conditions for gating, rather than making them explicit, decomposed training signals. Specifically, they fail to explicitly distinguish between uncertain retrieval and strong retrieval that conflicts with strong parametric beliefs. Inference-time monitoring does not necessarily reshape the underlying policy, and risk-control methods may be conservative in some cases—potentially abstaining even when retrieval evidence is strong, e.g., when internal beliefs and retrieved evidence disagree~\citep{mallen2023not,wen2024characterizing,jin2025disentangling}.

Our approach addresses this limitation by (i) separately estimating evidential support from parametric-only versus retrieval-augmented inputs, and (ii) converting their discrepancy into an explicit conflict signal. Rather than merely exposing the model to high-conflict cases during training, we use this explicit signal as a decomposed training signal to modulate alignment pressure, explicitly targeting disagreements between internal parametric beliefs and external evidence. More broadly, our method complements inference-time alignment or decoding-time interventions by shaping the training-time decision boundary for context utilization and abstention~\citep{hadji2024would}. This allows the model to effectively resolve conflicts by prioritizing reliable retrieval over misaligned priors. Consequently, our method operationalizes the critique that current incentives reward guessing, establishing a mechanism where abstention and context utilization are conditional on decomposed evidence and conflict resolution, rather than a single opaque confidence score.\looseness=-1

\cutsectionup
\section{PROBLEM FORMULATION}

\begin{figure}[t!]
    \centerline{
    \includegraphics[width=0.75\columnwidth]{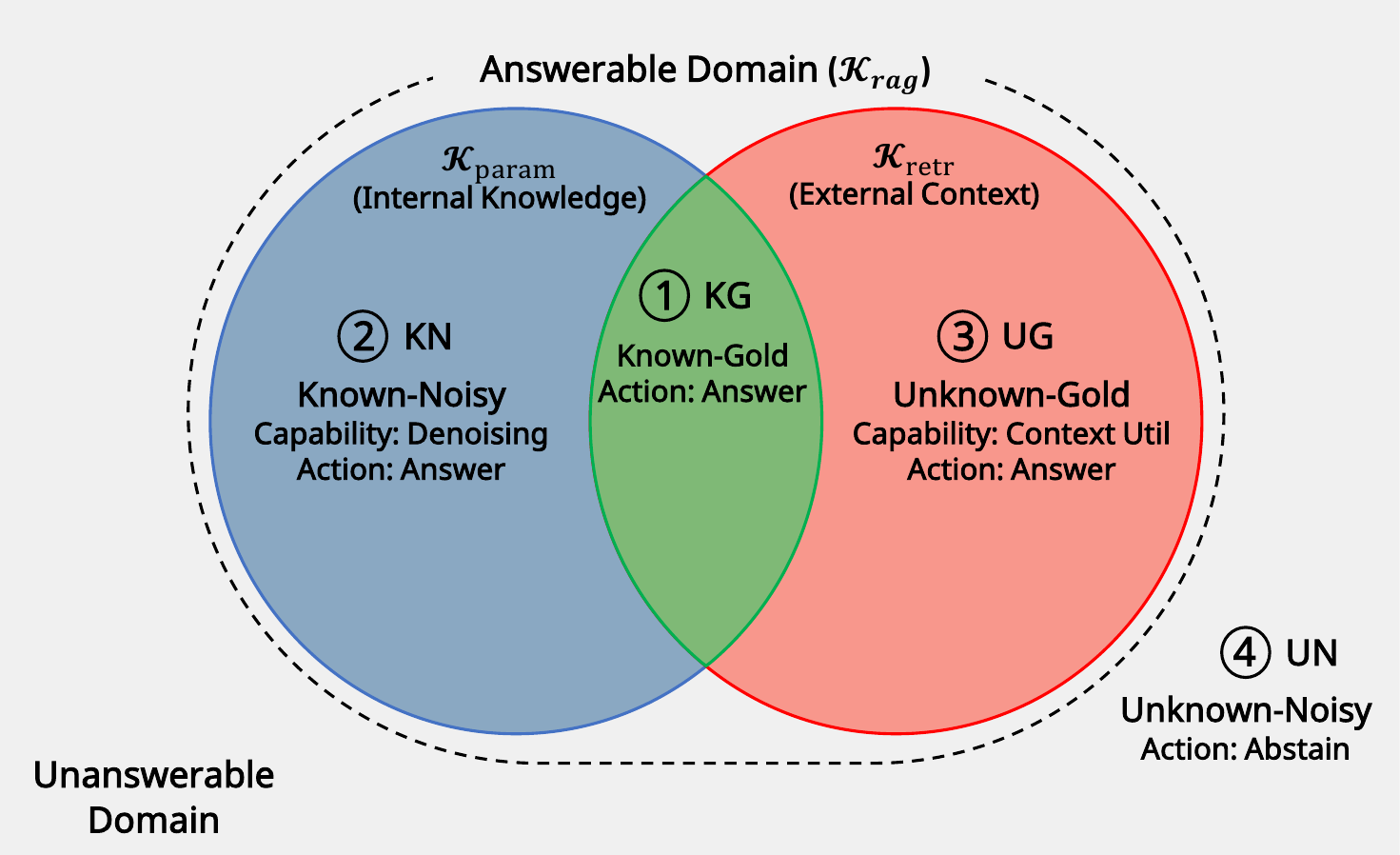}}
    \cutcaptionup
    \caption{Categorization of Knowledge Domains and Response Strategies in RAG Models.}
    \label{fig:quad}
    \vskip -0.7cm
\end{figure}

We formalize the problem of honest assessment, explicitly defining the knowledge boundaries, categorization criteria, and evaluation metrics employed in this study.

\cutsectionup
\subsection{Honest Assessment}
To rigorously evaluate the reliability of RAG systems, we first formalize the standard pipeline comprising a retriever $\mathcal{R}: \mathcal{Q} \to \mathcal{P}(\mathcal{D}_{\text{corpus}})$, which fetches relevant contexts $\mathcal{C}=\mathcal{R}(q) \subset \mathcal{D}_{\text{corpus}}$, and a generator $\mathcal{G}_\theta$. Let $\mathcal{Q}$ denote the query space and $\mathcal{A}$ the answer space. We define the generator $\mathcal{G}: \mathcal{Q} \times \mathcal{C} \rightarrow \mathcal{A}$ as a function producing an answer $y$ by modeling the conditional probability $P(y|q, \mathcal{C})$. Additionally, let $g: \mathcal{Q} \rightarrow \mathcal{A}$ be the ground truth function mapping a query to its correct response, $C(\cdot): \mathcal{Q}\times\mathcal{A}\to \{0,1\}$ be a correctness evaluation function and $CT(\cdot): \mathcal{C}\times\mathcal{A}\to\{0,1\}$ be a context-containment indicator that returns 1 if the context $c$ contains the ground-truth answer $g(q)$, and 0 otherwise.

\begin{table}[t!]
\centering
\caption{Summary of evaluation metrics.}
\label{tab:evaluation_metrics}
\small

\setlength{\tabcolsep}{4pt}
% [중요] X 컬럼의 세로 정렬을 'm'(middle)로 변경하여 텍스트가 수식 높이의 중앙에 오게 함
\renewcommand{\tabularxcolumn}[1]{m{#1}}
\cutcaptionup
\begin{tabularx}{\columnwidth}{@{} l c >{\raggedright\arraybackslash}X @{}}
\toprule
% 헤더를 \multicolumn{1}{c}{...}로 감싸서 강제로 가로 중앙 정렬
\multicolumn{1}{c}{\textbf{Metric}} & \multicolumn{1}{c}{\textbf{Formula}} & \multicolumn{1}{c}{\textbf{Description}} \\
\midrule

% --- Section 1 ---
\multicolumn{3}{@{}l}{\textit{\textbf{Answer Quality}}} \\
\addlinespace[3pt]
Recall & $\frac{|\mathcal{S}_{\text{ans}} \cap \mathcal{I}_{\text{ans}}|}{|\mathcal{I}_{\text{ans}}|}$ & Correct answers over all answerable queries \\
\addlinespace[5pt] % 수식 간 간격을 조금 더 주어 시원하게 보이게 함
Precision & $\frac{|\mathcal{S}_{\text{ans}} \cap \mathcal{I}_{\text{ans}}|}{|\mathcal{S}_{\text{ans}}|}$ & Correct answers over all attempted answers \\
\addlinespace[5pt]
F1 & $2 \cdot \frac{\text{Pre} \cdot \text{Rec}}{\text{Pre} + \text{Rec}}$ & Harmonic mean of Precision and Recall \\

\midrule

% --- Section 2 ---
\multicolumn{3}{@{}l}{\textit{\textbf{Retrieval Handling}}} \\
\addlinespace[3pt]
Denoise Rate & $\frac{|\mathcal{S}_{\text{ans}} \cap \mathcal{I}_{\text{noise}}|}{|\mathcal{I}_{\text{noise}}|}$ & Accuracy on queries with noisy context \\
\addlinespace[5pt]
Context Util. & $\frac{|\mathcal{S}_{\text{ans}} \cap \mathcal{I}_{\text{gold}}|}{|\mathcal{I}_{\text{gold}}|}$ & Accuracy on queries with golden context \\

\midrule

% --- Section 3 ---
\multicolumn{3}{@{}l}{\textit{\textbf{Abstain Quality}}} \\
\addlinespace[3pt]
Abstain Rec. & $\frac{|\mathcal{S}_{\text{abs}} \cap \mathcal{I}_{\text{unans}}|}{|\mathcal{I}_{\text{unans}}|}$ & Correct abstentions over unanswerable queries \\
\addlinespace[5pt]
Abstain Pre. & $\frac{|\mathcal{S}_{\text{abs}} \cap \mathcal{I}_{\text{unans}}|}{|\mathcal{S}_{\text{abs}}|}$ & Correct abstentions over all abstained queries \\
\addlinespace[5pt]
Abstain F1 & $2 \cdot \frac{\text{Pre} \cdot \text{Rec}}{\text{Pre} + \text{Rec}}$ & Harmonic mean of Abstain Precision and Recall \\

\midrule

% --- Section 4 ---
\textbf{Overall F1} & $2 \cdot \frac{\text{F1} \cdot \text{F1}_{\text{abs}}}{\text{F1} + \text{F1}_{\text{abs}}}$ & Trade-off between Answer and Abstain Quality \\

\bottomrule
\end{tabularx}
\end{table}

\mypara{Knowledge Components}
We posit that the system's total knowledge capability, denoted as $\mathcal{K}_{\text{rag}}$, is composed of two fundamental components: the parametric knowledge boundary of the LLM ($\mathcal{K}_{\text{param}}$) and the retrieval knowledge boundary ($\mathcal{K}_{\text{retr}}$). Formally, these are defined as:

\begin{align}
\mathcal{K}_{\text{param}} &= \{q \in \mathcal{Q} \mid C(\mathcal{G}(q, \emptyset)) = 1\} \\
\mathcal{K}_{\text{retr}} &= \{q \in \mathcal{Q} \mid \exists c \in \mathcal{R}(q) : CT(c,g(q))=1\,\}
% \text{Contains}(c, g(q))\}
\end{align}
% TODO

\noindent Consequently, the overall boundary is characterized as the union: $\mathcal{K}_{\text{rag}} = \mathcal{K}_{\text{param}} \cup \mathcal{K}_{\text{retr}}$. This formulation captures that a query is answerable if valid information resides within either the model's internal parameters or the retrieved external documents.

\begin{figure*}[t!]
    \centerline{
    \includegraphics[width=1.95\columnwidth]{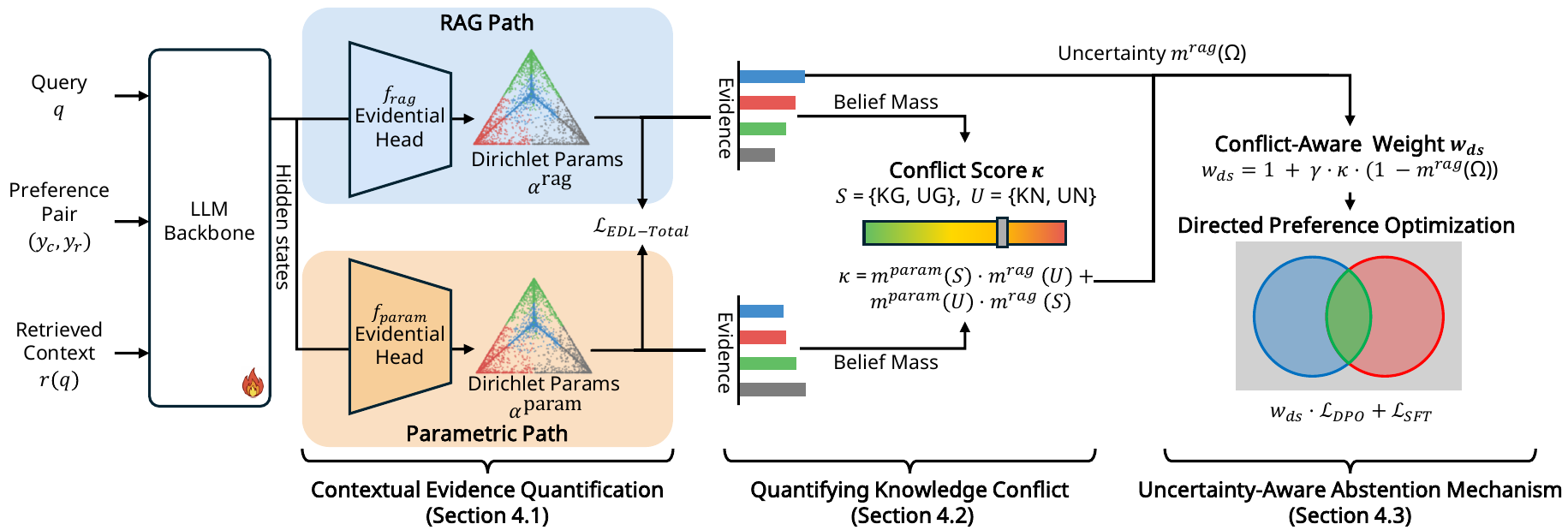}}
    \cutcaptionup
    \caption{Overview of the \model{}. The model consists of three  components: (1) Contextual Evidence Quantification, where evidential heads project representations from both RAG and Parametric paths into Dirichlet distributions ($\boldsymbol{\alpha}$); (2) Knowledge Conflict Quantification, which utilizes Dempster-Shafer Theory to fuse beliefs and compute a conflict score $\kappa$; and (3) Uncertainty-Aware Abstention Mechanism, which modulates the optimization objective based on the detected conflict to mitigate hallucinations.}
    \label{fig:model_overview}
\end{figure*}
% TODO

\mypara{Knowledge Quadrants and Ideal Alignment}
We categorize samples into four distinct knowledge quadrants based on the intersection of internal parametric knowledge ($\mathcal{K}_{\text{param}}$) and external retrieved evidence ($\mathcal{K}_{\text{retr}}$), as illustrated in Figure~\ref{fig:quad}. To ensure trustworthy system behavior, we define specific preference pairs—chosen ($y_c$) and rejected ($y_r$) responses—for each quadrant, strictly conditioning the model's operation on the composite knowledge boundary $\mathcal{K}_{\text{rag}}$.

First, we operationalize the boundaries as follows:
\begin{itemize}[leftmargin=*]
    \item \textbf{Determining $q \in \mathcal{K}_{\text{param}}$:} We sample $N$ answers $\{a_1, \dots, a_N\}$ from the model $\mathcal{G}(q, \emptyset)$ using different random seeds. If the proportion of correct answers (determined via lexical matching) exceeds a threshold $\delta$, we consider the knowledge internally available:
    \begin{equation}
        \frac{1}{N} \sum_{i=1}^{N} \mathbb{I}[C(a_i) = 1] > \delta
    \end{equation}
    \item \textbf{Determining $q \in \mathcal{K}_{\text{retr}}$:} We employ GPT-4o as a judge with a specialized prompt to assess whether the retrieved passages explicitly contain or imply the ground truth. If the context sufficiently supports the answer, we designate $q \in \mathcal{K}_{\text{retr}}$.
\end{itemize}

Based on these definitions, we construct the ideal alignment targets for each quadrant:

\begin{enumerate}[leftmargin=*,label=\large\protect\textcircled{\footnotesize\arabic*}]
    \item \textbf{Known-Gold (KG, $q \in \mathcal{K}_{\text{param}} \cap \mathcal{K}_{\text{retr}}$):} Both internal and external knowledge support the correct answer. Since the answer is well-supported, we designate the ground truth as the chosen response $y_c$. The rejected response $y_r$ is set to "I don't know" (IDK) to penalize unnecessary abstention.

    \item \textbf{Known-Noisy (KN, $q \in \mathcal{K}_{\text{param}} \setminus \mathcal{K}_{\text{retr}}$):} The model possesses the correct answer, but retrieval fails or provides noisy context. This tests \textit{Denoising} capability. Ideally, the model should prioritize internal knowledge. Thus, $y_c$ is the ground truth, while $y_r$ includes context-induced hallucinations and IDK responses to discourage vulnerability to noise and overly conservative behavior.

    \item \textbf{Unknown-Gold (UG, $q \in \mathcal{K}_{\text{retr}} \setminus \mathcal{K}_{\text{param}}$):} The model relies entirely on retrieved documents. This tests \textit{Context Utilization}. The ground truth serves as $y_c$. To ensure effective knowledge extraction, $y_r$ targets three failure modes: incorrect answers produced despite gold evidence in the provided context, incorrect answers produced without any context that reveal unsuppressed parametric hallucinations, and unjustified IDK responses.
    % failure to utilize the golden context, unsuppressed parametric hallucinations, % TODO

    \item \textbf{Unknown-Noisy (UN, $q \notin \mathcal{K}_{\text{param}} \cup \mathcal{K}_{\text{retr}}$):} Neither source contains the answer. Safety is paramount; the model must abstain. We designate IDK as $y_c$. Crucially, $y_r$ comprises parametric hallucinations, noise-derived answers, and even the \textit{ground truth} (to prevent unsupported "lucky guesses").
\end{enumerate}
% Insert Table here

\mypara{Evaluation Metrics}
Table ~\ref{tab:evaluation_metrics} outlines our evaluation framework.We define $\mathcal{S}_{\text{ans}}$ and $\mathcal{S}_{\text{abs}}$ as the sets of queries where the model decides to answer or abstain, respectively. To isolate specific performance behaviors, these prediction sets are intersected with ground-truth partitions, such as answerable queries ($\mathcal{I}_{\text{ans}} = \mathcal{K}_{\text{rag}}$), gold-only contexts ($\mathcal{I}_{\text{gold}} = \mathcal{K}_{\text{retr}} \setminus \mathcal{K}_{\text{param}}$), and noisy contexts ($\mathcal{I}_{\text{noise}} = \mathcal{K}_{\text{param}} \setminus \mathcal{K}_{\text{retr}}$).   We employ lexical matching to assess answer correctness, following existing work~\cite{sun2025divide}.  This allows precise quantification of the trade-off between utility (Answer Quality) and reliability (Abstain Quality) via the Overall F1 score. \looseness=-1

\section{Method}
Let $\mathcal{D} = \{(q, y_c, y_r, y_k)\}$ represent the preference alignment dataset, where each tuple contains a query $q$, a response preference pair $(y_c, y_r)$, and a one-hot encoded knowledge quadrant label $y_k$. We aim to optimize LLM $\mathcal{G}$ to satisfy the reliability constraints imposed by the preference pairs, while jointly learning to predict the corresponding label $y_k$.

Figure~\ref{fig:model_overview} illustrates the overall workflow. Our proposed framework, \model{}, first systematically aggregates evidence scores derived from retrieved contexts to evaluate the reliability and strength of external information (Section~\ref{sec:evidence_score}). Subsequently, it computes a conflict score to quantitatively measure the discrepancy between the model's intrinsic parametric knowledge and the provided external evidence (Section~\ref{sec:conflict_score}). Guided by this metric, the model dynamically modulates the preference alignment objective to prioritize either robustness or context utilization, thereby ensuring a well-calibrated mapping to the target knowledge quadrant (Section~\ref{sec:calibrate}). These techniques effectively mitigate contextual noise and calibrate the model's behavioral alignment. In the following section, we describe the proposed framework in detail. 

\subsection{Contextual Evidence Quantification}
\label{sec:evidence_score}
Standard models typically optimize the conventional cross-entropy loss, which compels them to produce confident point estimates, often leading to uncalibrated overconfidence and conflated epistemic and aleatoric uncertainties. We adopt Evidential Deep Learning (EDL) to disentangle these uncertainties by capturing the predictive distribution density~\citep{sensoy2018evidential}. Moreover, since a single uncertainty estimate in RAG obscures whether uncertainty arises from intrinsic ignorance or knowledge conflicts, we extend EDL with a dual-head architecture to independently quantify evidential support from parametric memory and retrieved context.

\mypara{Evidential Formulation via Subjective Logic} 
To distinctively capture the uncertainty from external and internal sources, we integrate two auxiliary evidential heads into the LLM $\mathcal{G}$: $f_{\text{rag}}$ and $f_{\text{param}}$. These heads take the hidden states from the retrieval-augmented input and the parametric-only input, respectively, projecting them into logit vectors $z \in \mathbb{R}^4$ corresponding to the four knowledge quadrants. The parametric-only input is formed using the same query without any retrieval context, allowing $f_{\text{param}}$ to estimate an evidential prior grounded in parametric memory even when no documents are provided. This prior is defined in the same quadrant-based evidential space as $f_{\text{rag}}$, which enables consistent internal–external comparison and calibration when retrieved evidence is provided.

To disentangle different types of uncertainty, we parameterize the categorical distribution over the quadrants using a Dirichlet prior. This formulation, rooted in Subjective Logic, allows the framework to decompose collected evidence into belief masses for each quadrant and a comprehensive uncertainty term~\cite{jsang2018subjective}. The logit vector $z$ from each head is transformed into non-negative evidence $e \in \mathbb{R}^4$ via the Softplus activation to ensure smooth optimization:
\begin{align}
e_k = \text{Softplus}(z_k), \quad k \in \{1, \dots, K\}
\end{align}
According to Subjective Logic, the evidence $e_k$ is related to the belief mass $b_k$ and the uncertainty mass $u$ as follows:
\begin{align}
b_k = \frac{e_k}{S}, \quad u = \frac{K}{S}, \quad \text{where } S = \sum_{k=1}^{K} (e_k + 1)
\end{align}
Here, $S$ represents the Dirichlet strength. The sum of all masses is constant: $\sum_{k=1}^K b_k + u = 1$. The corresponding Dirichlet distribution parameters are derived as $\alpha_k = e_k + 1$. Intuitively, if the model collects more evidence for a specific quadrant, the corresponding belief mass $b_k$ increases, and the global uncertainty $u$ decreases.

\mypara{Optimization Objective} To calibrate these evidential outputs, we employ a hybrid objective function that combines the expected risk minimization with a regularizer. Our framework optimizes the objective by integrating cross-entropy loss over the Dirichlet-defined simplex  $\text{Dir}(\boldsymbol{p} | \boldsymbol{\alpha})$, thereby minimizing the overall Bayes risk. This yields the analytical form involving the digamma function $\psi(\cdot)$:\looseness=-1
\begin{align}
\mathcal{L}_{\text{fit}}(\boldsymbol{\alpha}, \boldsymbol{y}) = \sum_{k=1}^K y_k \left( \psi(S) - \psi(\alpha_k) \right)
\end{align}
where $\boldsymbol{y}$ is one-hot target vector. This term encourages the predictive distribution to concentrate around the ground truth quadrant.

While $\mathcal{L}_{\text{fit}}$ encourages correct classification, it does not guarantee that evidence for incorrect classes is suppressed. To mitigate this, we incorporate a Kullback-Leibler (KL) divergence that penalizes the divergence between the predicted posterior and a uniform Dirichlet prior $\text{Dir}(\mathbf{1})$, which represents total uncertainty ($u=1$).
\begin{align}
\mathcal{L}_{\text{KL}}(\boldsymbol{\alpha}, \boldsymbol{y}) = \text{KL}\left[ \text{Dir}(\tilde{\boldsymbol{\alpha}}) \| \text{Dir}(\mathbf{1}) \right]
\end{align}
Here, $\tilde{\boldsymbol{\alpha}}=\boldsymbol{y}(1-\boldsymbol{y})\!\odot\!\boldsymbol{\alpha}$ represents the adjusted parameters used to penalize misleading evidence assigned to non-target classes.

The individual evidential loss for a given head is computed as a weighted sum, where $\lambda_t$ is an annealing coefficient that linearly increases over training steps $t$ to prevent premature convergence:
\begin{align}
\mathcal{L}_{\text{EDL}}(\boldsymbol{\alpha}, \boldsymbol{y}) = \mathcal{L}_{\text{fit}}(\boldsymbol{\alpha}, \boldsymbol{y}) + \lambda_t \mathcal{L}_{\text{KL}}(\boldsymbol{\alpha}, \boldsymbol{y})
\end{align}
Finally, since our framework relies on two distinct evidential sources to detect knowledge conflicts, we apply this objective to the outputs of both heads. Let $\boldsymbol{\alpha}^{\text{rag}}$ and $\boldsymbol{\alpha}^{\text{param}}$ denote the Dirichlet parameters derived from $f_{\text{rag}}$ and $f_{\text{param}}$, respectively. The total evidential learning objective is defined as the sum of losses for both contexts:
\begin{align}
\mathcal{L}_{\text{EDL-Total}} = \mathcal{L}_{\text{EDL}}(\boldsymbol{\alpha}^{\text{rag}}, \boldsymbol{y}) + \mathcal{L}_{\text{EDL}}(\boldsymbol{\alpha}^{\text{param}}, \boldsymbol{y})
\end{align}

\subsection{Quantifying Knowledge Conflict}
\label{sec:conflict_score}
Having independently quantified the evidential distributions from parametric and retrieval sources, we next assess their alignment. Detecting discrepancies where the model's internal priors contradict retrieved evidence is critical, as such conflicts are often precursors to hallucinations. To rigorously measure this discord, we employ Dempster-Shafer Theory (DST) to fuse the distinct evidential sources. This fusion process yields a conflict score $\kappa$, which serves as a robust proxy for identifying hallucinations and delineating the model's knowledge boundaries. 

\mypara{Belief Mass Reduction} While our evidential heads predict fine-grained knowledge quadrants, the fundamental conflict in RAG scenarios stems from the validity of the evidence: whether the source provides sufficient information to support an answer or fails to do so (due to noise or ignorance). To capture this from a RAG perspective, we aggregate the quadrant evidence into two high-level hypotheses: Supported ($\mathcal{S}$) and Unsupported ($\mathcal{U}$).

\begin{align}
\mathcal{S} = \{\text{KG, UG}\}, \quad \mathcal{U} = \{\text{KN, UN}\}
\end{align}

For each source $s \in \{\text{rag, param}\}$, corresponding to the evidential outputs of heads $f_{\text{rag}}$ and $f_{\text{param}}$ respectively, we reduce the 4-dimensional Dirichlet parameters into a 2-dimensional belief mass assignment. The mass functions for these hypotheses and the global uncertainty ($\Omega$) are derived by summing the belief masses of the constituent quadrants: 
\begin{align}
m^s(\mathcal{S}) = \sum_{k \in \mathcal{S}} b_k^s, \quad m^s(\mathcal{U}) = \sum_{k \in \mathcal{U}} b_k^s, \quad m^s(\Omega) = u^s
\end{align}

This reduction allows us to focus on the semantic validity of the evidence—specifically, whether the parametric memory and the retrieved context support the generation of an answer or remain unsupported—while strictly preserving the global uncertainty $m^s(\Omega)$.\looseness=-1

\mypara{Conflict Score ($\kappa$) Calculation} We utilize Dempster’s Rule of Combination to rigorously quantify the tension between the model's internal beliefs ($m^{\text{param}}$) and the external evidence ($m^{\text{rag}}$). The conflict score $\kappa$ represents the probability mass assigned to the empty set, arising when sources support mutually exclusive hypotheses:\begin{equation}\kappa = m^{\text{param}}(\mathcal{S}) \cdot m^{\text{rag}}(\mathcal{U}) + m^{\text{param}}(\mathcal{U}) \cdot m^{\text{rag}}(\mathcal{S})\end{equation}Crucially, this formulation inherently accounts for uncertainty. Since the sum of masses is fixed to 1, a high uncertainty $m(\Omega)$ in either source naturally reduces the masses assigned to $\mathcal{S}$ and $\mathcal{U}$, thereby suppressing $\kappa$. Thus, a high $\kappa$ value indicates a critical dissonance where one source strongly supports the answer while the other strongly refutes it.

\subsection{Uncertainty-Aware Abstention Mechanism}\label{sec:calibrate}The ultimate goal of our framework is to calibrate the model's behavior so that it selectively abstains when intrinsic knowledge is insufficient or contradicts reliable external evidence. Standard preference optimization (e.g., DPO) treats all training samples uniformly, which is suboptimal for learning these delicate decision boundaries~\cite{rafailov2023direct}. To address this, we propose a comprehensive training strategy that prioritizes learning from high-conflict, high-evidence scenarios.\looseness=-1

\mypara{Conflict-Aware Loss Modulation} We hypothesize that the most critical signal for learning robustness lies in samples where the model's internal belief conflicts with the retrieved context—but only if that context is trustworthy. We formulate a dynamic weighting factor $w_{\text{ds}}$ applied to the alignment loss:

\begin{align}w_{\text{ds}} = 1 + \gamma \cdot \kappa \cdot (1 - m^{\text{rag}}(\Omega))
\label{eq:w_ds}
\end{align}

Here, $\gamma$ is a hyperparameter controlling the sensitivity of the modulation, and the term $(1 - m^{\text{rag}}(\Omega))$ acts as a reliability gate. Specifically, when the conflict $\kappa$ is high and the retrieval uncertainty is low, the weight $w_{\text{ds}}$ increases, forcing the model to focus on resolving the dissonance—typically by aligning with strong external evidence or correcting hallucinations. Conversely, when the retrieval itself is uncertain ($m^{\text{rag}}(\Omega) \approx 1$), this gate effectively closes, keeping $w_{\text{ds}}$ near $1.0$. This mechanism prevents the model from aggressively updating its policy based on ambiguous or noisy retrieval results.

\mypara{Preference Alignment}
To align the model with human preferences, we employ Direct Preference Optimization (DPO), which optimizes the policy directly using pairs of chosen ($y_c$) and rejected ($y_r$) responses without requiring a separate reward model. Given a query $q$ and retrieved context $r(q)$, the DPO loss is formulated as:
\begin{equation}
\resizebox{1.0\hsize}{!}{$
    \mathcal{L}_{\text{DPO}}(\pi_\theta; \pi_{\text{ref}}) = -\log \sigma \left( \tau \left[ \log \frac{\pi_\theta(y_c|q, r(q))}{\pi_{\text{ref}}(y_c|q, r(q))} - \log \frac{\pi_\theta(y_r|q, r(q))}{\pi_{\text{ref}}(y_r|q, r(q))} \right] \right),
$}
\vspace{2mm}
\end{equation}
where $\pi_\theta$ is the policy being trained, $\pi_{\text{ref}}$ is the frozen reference model, and $\tau$ is the temperature parameter. 

However, DPO primarily focuses on degrading the likelihood of rejected responses rather than actively improving the chosen ones. To mitigate this and maintain generation coherence, we add a supervised fine-tuning (SFT) auxiliary loss on the chosen responses:
\begin{equation}
    \mathcal{L}_{\text{SFT}}(\pi_\theta) = - \frac{1}{T} \sum_{t=1}^{T} \log \pi_\theta(y_{c,t} | q, r(q), y_{c, <t}),
\end{equation}
where $y_{c,t}$ denotes the $t$-th token of the chosen response. This auxiliary term explicitly reinforces the model's ability to generate high-quality preferred outputs.

\mypara{Unified Training Objective} We integrate the evidential learning and the preference alignment into a unified objective. The final total loss $\mathcal{L}_{\text{total}}$ is a weighted combination of the conflict-modulated DPO loss, the total evidential loss, and the SFT auxiliary loss:\begin{align}\mathcal{L}_{\text{total}} = w_{\text{ds}} \cdot \mathcal{L}_{\text{DPO}}(\pi\theta; \pi_{\text{ref}}) + \mathcal{L}_{\text{EDL-Total}} + \mathcal{L}_{\text{SFT}}\end{align}

\noindent where $w_{\text{ds}}$ is the conflict-aware modulation factor derived in~\ref{eq:w_ds}.

% 색상 정의
\definecolor{grayrow}{gray}{0.92}
\definecolor{oursrow}{RGB}{235, 242, 250} % 연한 파란색 (제안 모델 강조용)

\begin{table*}[t!]
\centering
\caption{Performance comparison across three QA datasets using Qwen3-8B and Llama3-8B backbones. Metrics are categorized into: \textbf{Answer Quality}, \textbf{Retrieval Handling}, and \textbf{Abstention Quality}.
\textbf{Overall F1} represents the harmonic mean balancing answer generation and abstention.}
\label{tab:performance_comparison}

% 스타일 설정
\setlength{\aboverulesep}{0pt}
\setlength{\belowrulesep}{0pt}
\renewcommand{\arraystretch}{1.2} 
\setlength{\tabcolsep}{4.5pt} % 열 간격 미세 조정

% 열 구조: Method | Gen(3) | Ret(2) | Abs(3) | Overall(1)
\cutcaptionup
\begin{tabular}{c | ccc | cc | ccc | c}
\toprule
\multirow{2.5}{*}{\textbf{Method}} & \multicolumn{3}{c|}{\textbf{Answer Quality}} & \multicolumn{2}{c|}{\textbf{Retrieval Handling}} & \multicolumn{3}{c|}{\textbf{Abstention Quality}} & \multirow{2.5}{*}{\textbf{Overall F1}} \\
\cmidrule(lr){2-4} \cmidrule(lr){5-6} \cmidrule(lr){7-9}
 & Recall & Prec. & F1 & Denoise & Context & Recall & Prec. & F1 & \\
\midrule

% ================= Qwen3-8B =================
\rowcolor{grayrow} \multicolumn{10}{l}{\textbf{Qwen3-8B-Base}} \\
\hspace{3mm}Logprob ($t=-2$)      & 0.308 & 0.933 & 0.463 & 0.303 & 0.221 & 0.050 & 0.564 & 0.091 & 0.153 \\
\hspace{3mm}Logprob ($t=-1$)      & 0.197 & 0.933 & 0.325 & 0.187 & 0.144 & 0.689 & 0.360 & 0.473 & 0.385 \\
\hspace{3mm}ICL                 & 0.178 & 0.130 & 0.150 & 0.184 & 0.142 & 0.048 & 0.404 & 0.085 & 0.109 \\
\hspace{3mm}Self-Consistency         & 0.203 & 0.415 & 0.273 & 0.219 & 0.142 & 0.777 & 0.350 & 0.483 & 0.349 \\
\hspace{3mm}$p_{\text{true}}$ ($t=0.5$) & 0.209 & 0.931 & 0.342 & 0.245 & 0.128 & 0.521 & 0.399 & 0.452 & 0.389 \\
\hspace{3mm}$p_{\text{true}}$ ($t=0.7$) & 0.133 & 0.973 & 0.234 & 0.161 & 0.036 & 0.883 & 0.353 & 0.504 & 0.320 \\
\hspace{3mm}$p_{\text{true}}$ ($t=0.9$) & 0.014 & \textbf{1.000} & 0.029 & 0.014 & 0.001 & \textbf{0.990} & 0.294 & 0.453 & 0.054 \\
\hspace{3mm}DTA                 & 0.436 & 0.410 & 0.422 & 0.380 & 0.319 & 0.476 & 0.566 & 0.517 & 0.465 \\ 
% 제안 모델 행 (Qwen3)
\rowcolor{oursrow} 
\hspace{3mm}\textbf{\model{}}   & \textbf{0.622} & 0.532 & \textbf{0.573} & \textbf{0.606} & \textbf{0.454} & 0.408 & \textbf{0.715} & \textbf{0.520} & \textbf{0.545} \\

\midrule

% ================= Llama3-8B =================
\rowcolor{grayrow} \multicolumn{10}{l}{\textbf{Llama3.1-8B}} \\
\hspace{3mm}Logprob ($t=-2$)      & 0.509 & 0.910 & \textbf{0.653} & \textbf{0.568} & \textbf{0.392} & 0.314 & 0.424 & 0.361 & 0.465 \\
\hspace{3mm}Logprob ($t=-1$)      & 0.282 & 0.914 & 0.431 & 0.346 & 0.199 & 0.775 & 0.340 & 0.472 & 0.451 \\
\hspace{3mm}ICL                 & 0.474 & 0.336 & 0.394 & 0.548 & 0.326 & 0.014 & \textbf{0.800} & 0.027 & 0.050 \\
\hspace{3mm}Self-Consistency         & 0.386 & 0.492 & 0.432 & 0.392 & 0.300 & 0.559 & 0.368 & 0.444 & 0.438 \\
\hspace{3mm}$p_{\text{true}}$ ($t=0.5$) & 0.364 & 0.931 & 0.523 & 0.375 & 0.235 & 0.311 & 0.556 & 0.268 & 0.312 \\
\hspace{3mm}$p_{\text{true}}$ ($t=0.7$) & 0.337 & 0.934 & 0.495 & 0.352 & 0.207 & 0.497 & 0.484 & 0.293 & 0.368 \\
\hspace{3mm}$p_{\text{true}}$ ($t=0.9$) & 0.253 & \textbf{0.952} & 0.400 & 0.300 & 0.130 & \textbf{0.779} & 0.402 & 0.222 & 0.255 \\
\hspace{3mm}DTA                 & 0.490 & 0.544 & 0.516 & 0.510 & 0.322 & 0.596 & 0.481 & 0.533 & 0.524 \\ 
% 제안 모델 행 (Llama3)
\rowcolor{oursrow} 
\hspace{3mm}\textbf{\model{}}   & \textbf{0.537} & 0.680 & 0.600 & 0.504 & 0.367 & 0.759 & 0.506 & \textbf{0.607} & \textbf{0.604} \\
\bottomrule
\end{tabular}

\end{table*}

Optimizing this objective, \model{} jointly learns to adhere to human preferences while simultaneously calibrating its predictive distributions to accurately reflect knowledge boundaries and uncertainty.\looseness=-1

\section{Experiment}
We evaluate \model{} on multiple RAG datasets and perform various analyses to assess how our framework improves calibration quality. Due to space constraints, further details, including implementation details and results from varying LLMs, are provided in the Appendix.\looseness=-1

\subsection{Performance Evaluation}

\mypara{Datasets} Consistent with prior RAG studies~\cite{fang2024enhancing,sun2025divide}, we validate the effectiveness of our proposed framework across diverse question-answering scenarios by conducting comprehensive evaluations on three widely established open-domain QA benchmarks: Natural Questions (NQ)~\cite{kwiatkowski2019natural}, TriviaQA~\cite{joshi2017triviaqa}, and WebQuestions (WebQ)~\cite{berant2013semantic}. These datasets were selected to cover a broad spectrum of query types, ranging from real-world user search queries (NQ) to complex trivia-based reasoning (TriviaQA) and entity-centric questions (WebQ). For retrieval, we utilize Dense Passage Retrieval (DPR)~\cite{karpukhin2020dense} to obtain the top-3 passages from Wikipedia, assessing knowledge availability by verifying the presence of the ground truth answer within the retrieved contexts.

To evaluate generalization capabilities on information unseen during pre-training, we construct the Wiki Event benchmark by parsing the 'Events' sections of Wikipedia annual pages ranging from 2010 to 2025. This temporal coverage specifically targets events post-dating standard knowledge cutoffs. For specific event descriptions, we scrape the referenced external news articles to establish them as the gold contexts. We then leverage GPT-4o to generate factual question-answer pairs derived directly from these Wikipedia summaries. Crucially, we also synthesize counterfactual contexts by systematically perturbing key entities in the gold contexts to simulate realistic scenarios of noisy or conflicting retrieval.\looseness=-1

\mypara{Baselines} We implement five calibration strategies as baselines: (1) Logprob calculates the average log-probability of all tokens in the generated answer, serving as a basic confidence metric~\cite{kadavath2022language}. (2) ICL prompts the model to verbally self-evaluate the correctness of its answer~\cite{wei2022chain}. (3) Self-Consistency computes confidence based on the agreement rate among multiple sampled outputs for the same query~\cite{wangself}. (4) $P_{\text{true}}$ extracts the specific softmax probability assigned to the "True" token given a verification prompt, providing a continuous score~\cite{kadavath2022language}. (5) DTA employs DPO to train models to abstain from answering when retrieved evidence is insufficient, using curated abstention datasets~\cite{sun2025divide}.

\mypara{Results} Table~\ref{tab:performance_comparison} presents a performance comparison of calibration strategies regarding Answer Quality, Retrieval Handling, and Abstention Quality. \model{} achieves the highest Overall F1 scores across both Llama3-8B~\cite{dubey2024llama} and Qwen3-8B-Base~\cite{yang2025qwen3} backbones. While confidence based methods (e.g., Logprob, $P_{\text{true}}$) falter due to threshold sensitivity and inability to distinguish uncertainty types, and supervised baselines like DTA are prone to memorizing specific refusal patterns rather than learning robust decision boundaries, our evidential framework overcomes these limitations. By explicitly quantifying belief masses to identify "true unknowns" (epistemic uncertainty), \model{} delivers superior Abstention Precision, effectively balancing answer generation and abstention without succumbing to the overfitting or calibration pitfalls observed in prior baselines.  

\begin{figure}[!t] 
        \centering \includegraphics[width=0.95\linewidth]{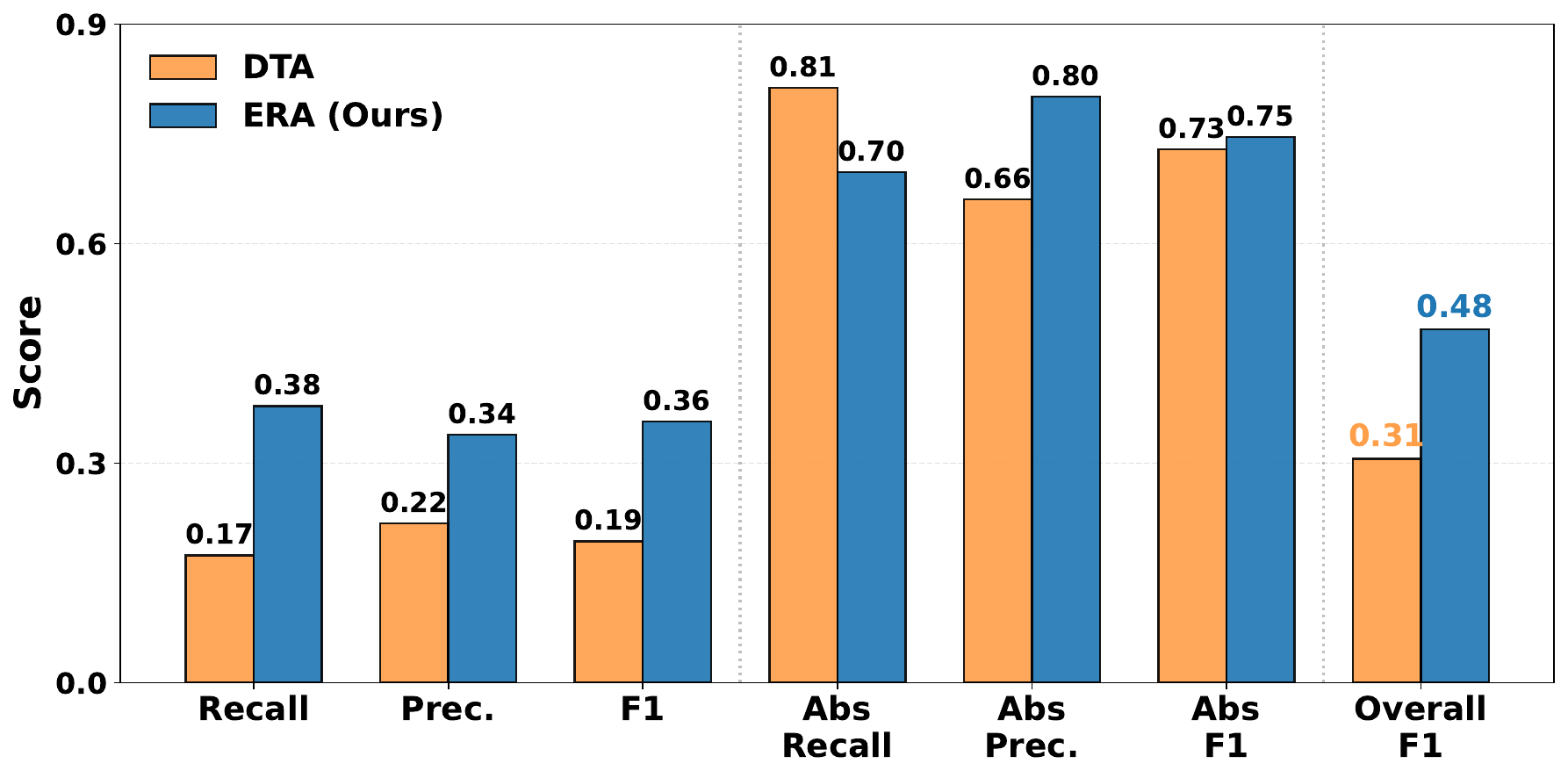}
        \cutcaptionup
        \caption{Generalizability evaluation on the Wiki Event dataset using Qwen3-8B-Base.}
    \label{fig:wiki_result}
\end{figure}

Figure~\ref{fig:wiki_result} demonstrates that \model{} significantly outperforms the strongest baseline (DTA) on the out-of-distribution Wiki Event dataset, achieving a superior Overall F1. While DTA achieves high Abstention Recall, its critically low Answer F1 exposes a vulnerability to overfitting: the model appears to memorize specific refusal patterns from the training data, leading to indiscriminate over-abstention on unseen events. In contrast, \model{} effectively mitigates this overfitting by relying on disentangled uncertainty estimates, doubling the Answer F1 while maintaining high Abstention F1, thereby validating its robust generalization capability.

\begin{figure*}[t!]
\centering
\begin{subfigure}[t!]{0.32\textwidth}
\captionsetup{width=.99\columnwidth}
\includegraphics[width=0.99\columnwidth]{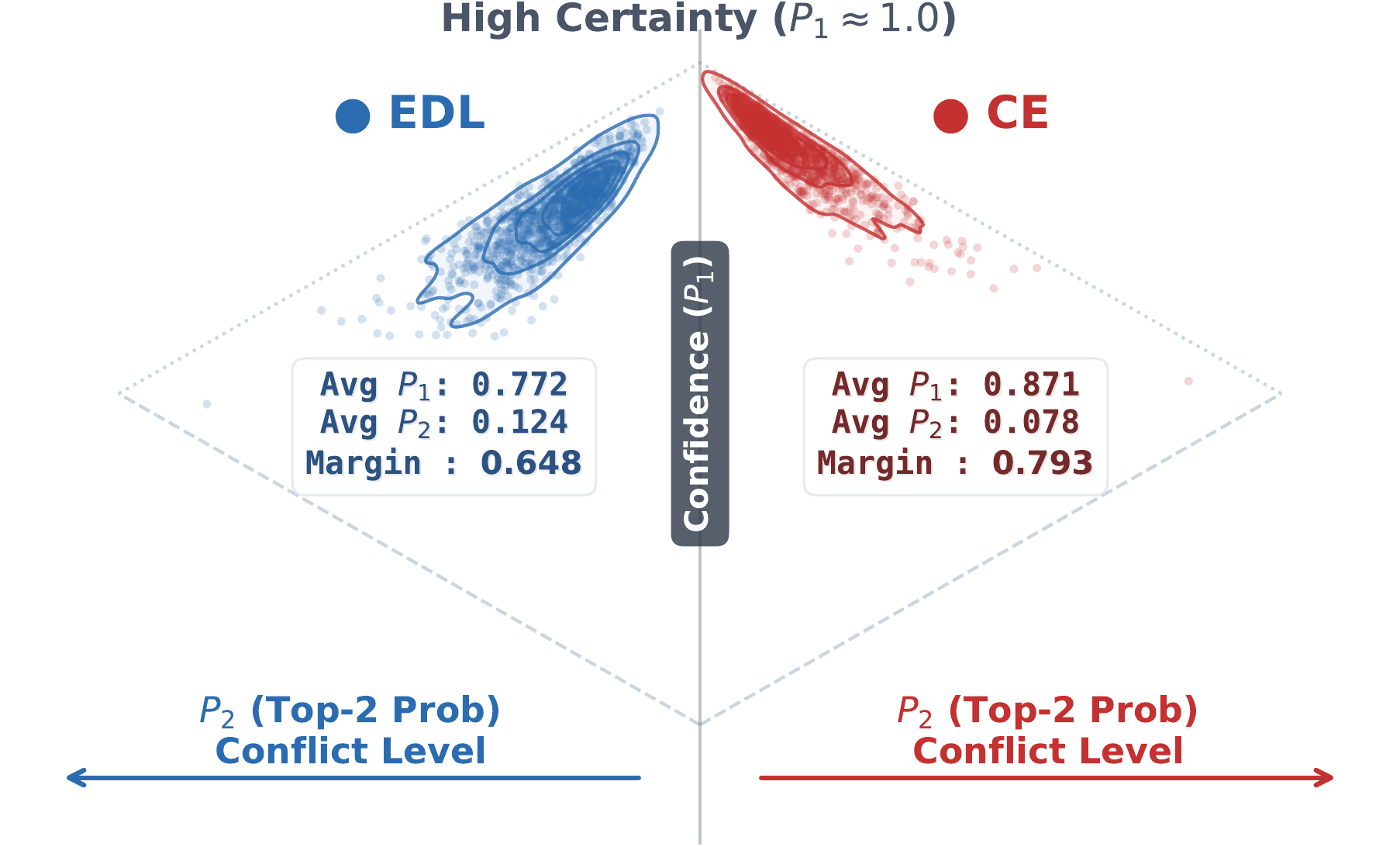}

    \cutcaptionup
    \caption{Comparison of Uncertainty Distributions}
    \label{fig:uncertainty}
\end{subfigure}
\hspace{1mm}
\begin{subfigure}[t!]{0.33\textwidth}
\captionsetup{width=.99\linewidth}
\includegraphics[width=0.99\columnwidth]{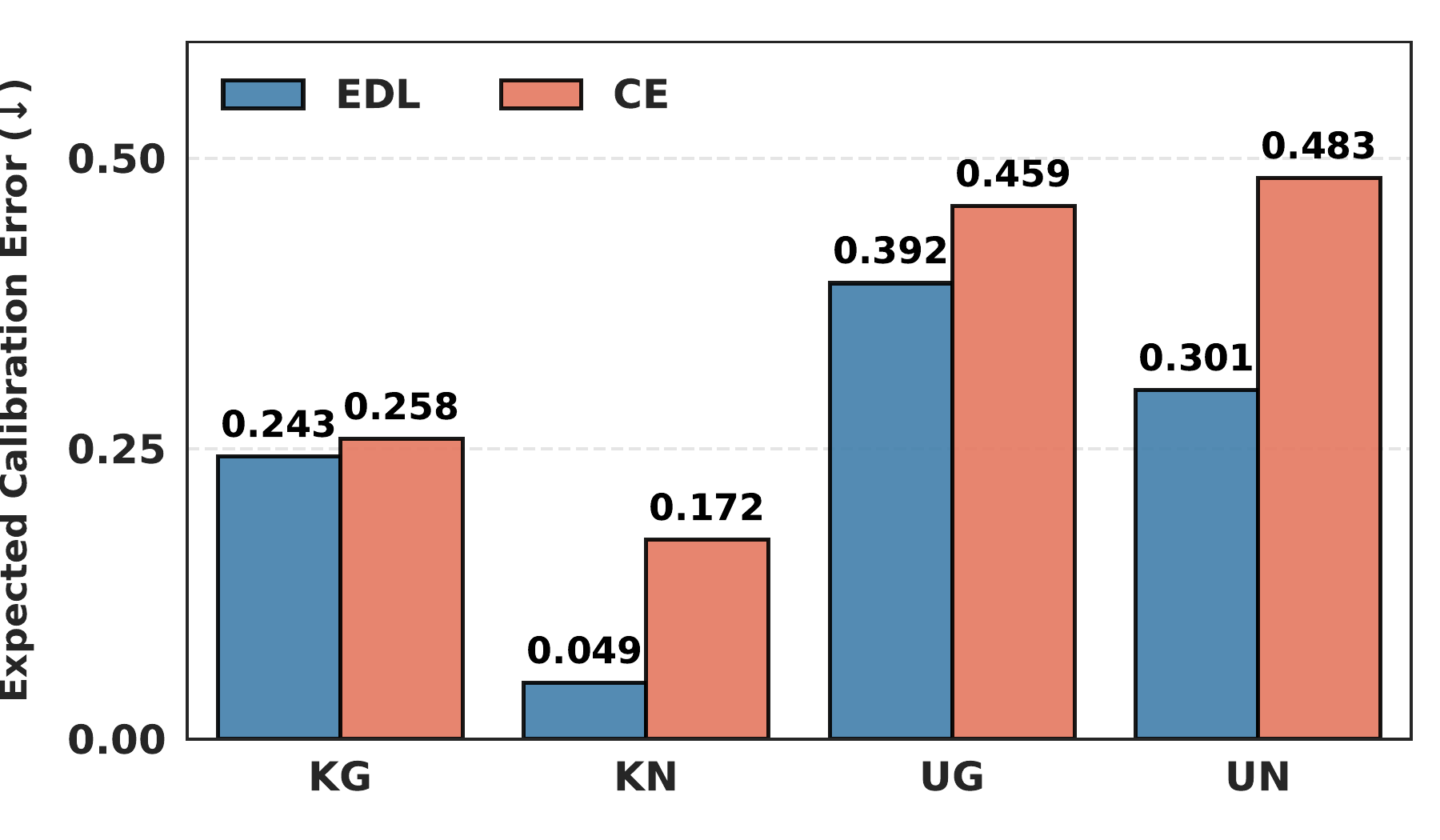}
    \cutcaptionup
    \caption{Comparison of Expected Calibration Error}
    \label{fig:ece_error}
\end{subfigure}
\begin{subfigure}[t!]{0.33\textwidth}
\captionsetup{width=.99\linewidth}
\includegraphics[width=0.99\columnwidth]{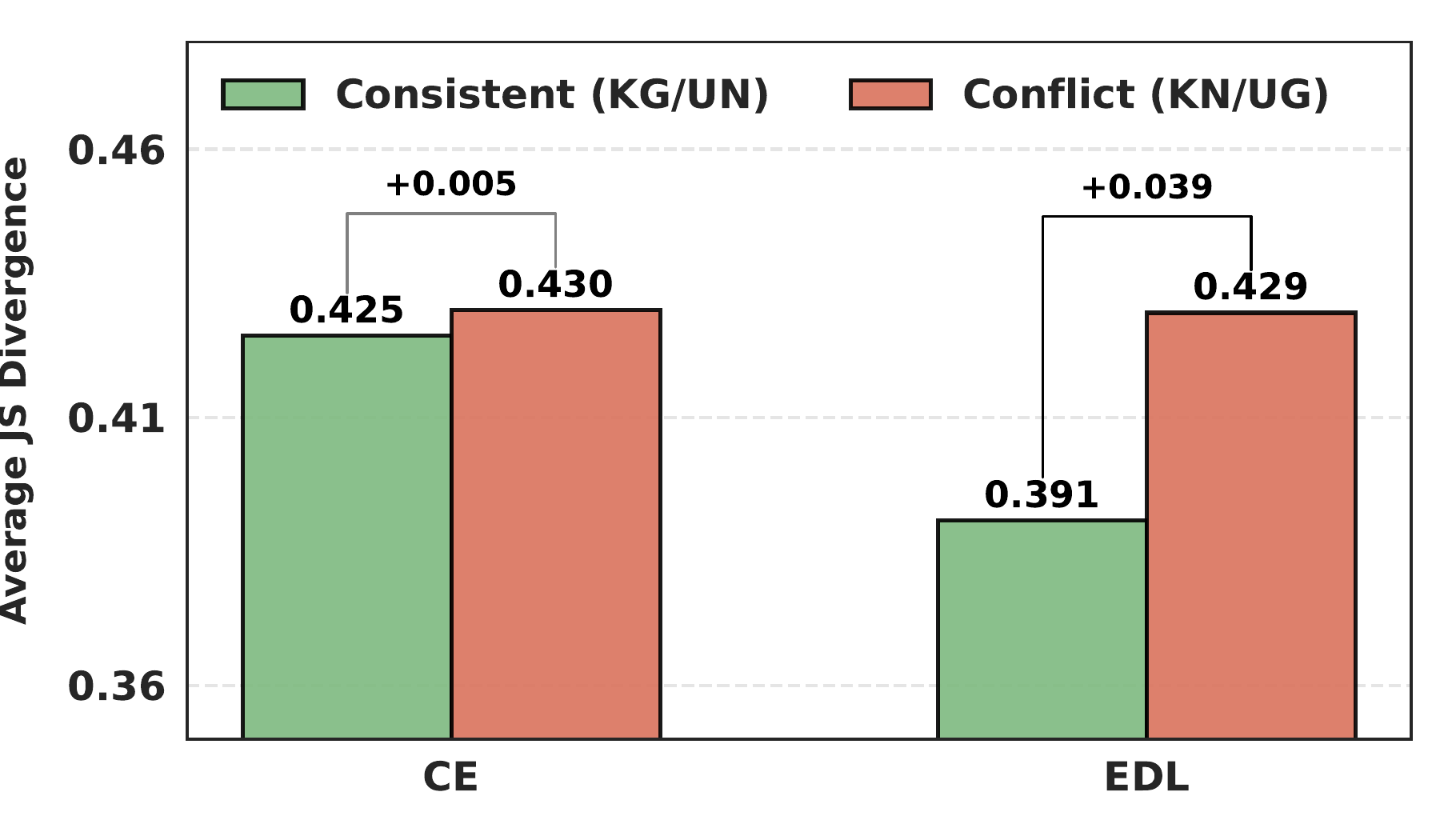}
    \cutcaptionup
    \caption{Conflict Sensitivity Analysis}
    \label{fig:js_div}
\end{subfigure}
\cutcaptionup
\caption{Quantitative analysis of uncertainty. This figure compares the proposed Evidential Deep Learning (EDL) model against the Cross-Entropy (CE) baseline across three dimensions: (a) A visualization of prediction probability distributions in conflict situations (KN, UG); (b) A comparison of Expected Calibration Error (ECE) across data types; and (c) An evaluation of model sensitivity via the Jensen-Shannon Divergence (JSD) between consistent and conflicting information.\looseness=-1}
\end{figure*}

\begin{figure}[!t]
    \centering
    \begin{subfigure}[b]{.49\linewidth}
        \centering\captionsetup{width=1.0\linewidth}
        \includegraphics[width=\linewidth]{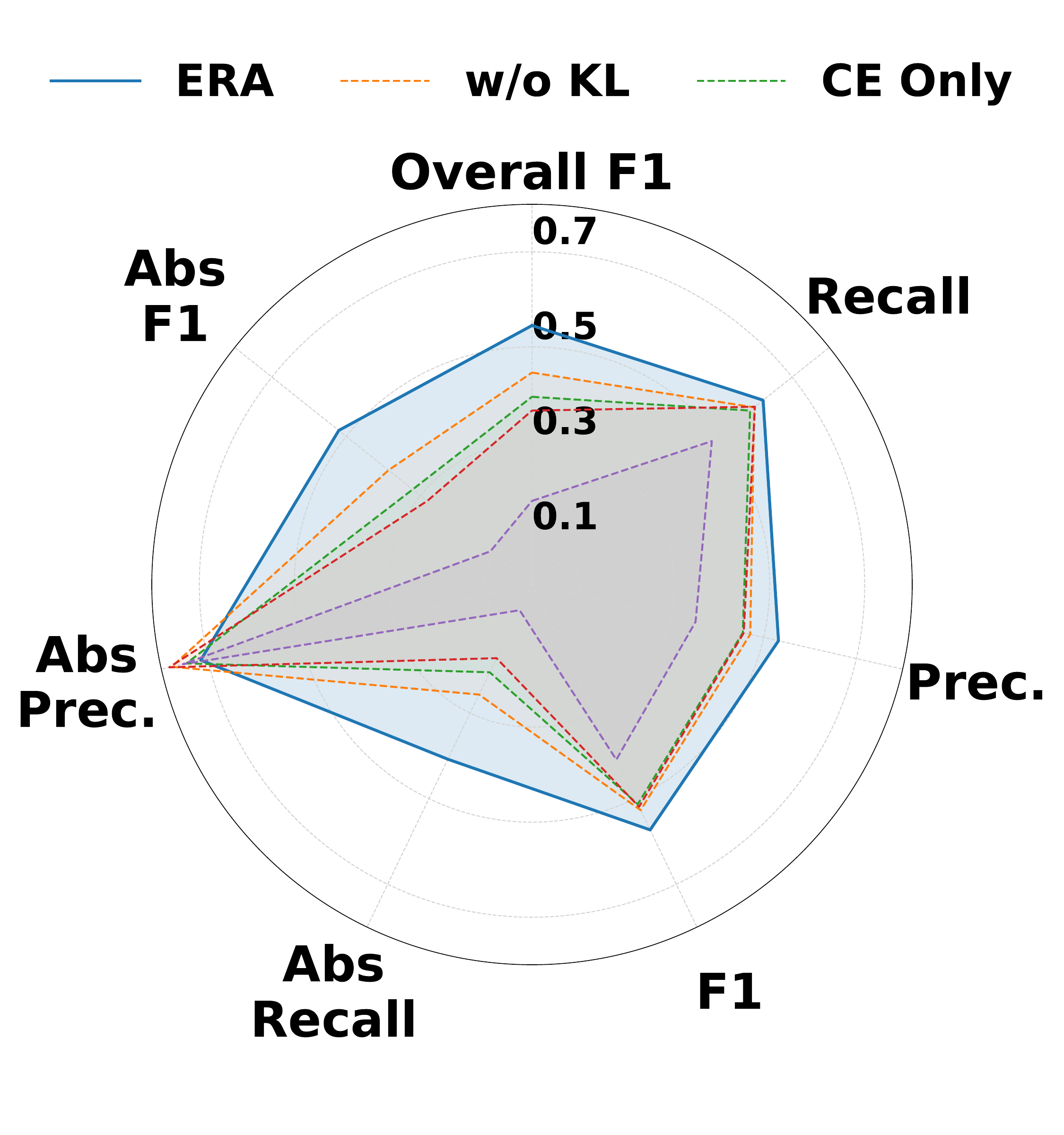}
        \caption{Qwen3-8B-Base}
    \end{subfigure}
    \begin{subfigure}[b]{.49\linewidth}
        \centering\captionsetup{width=1.0\linewidth}
        \includegraphics[width=\linewidth]{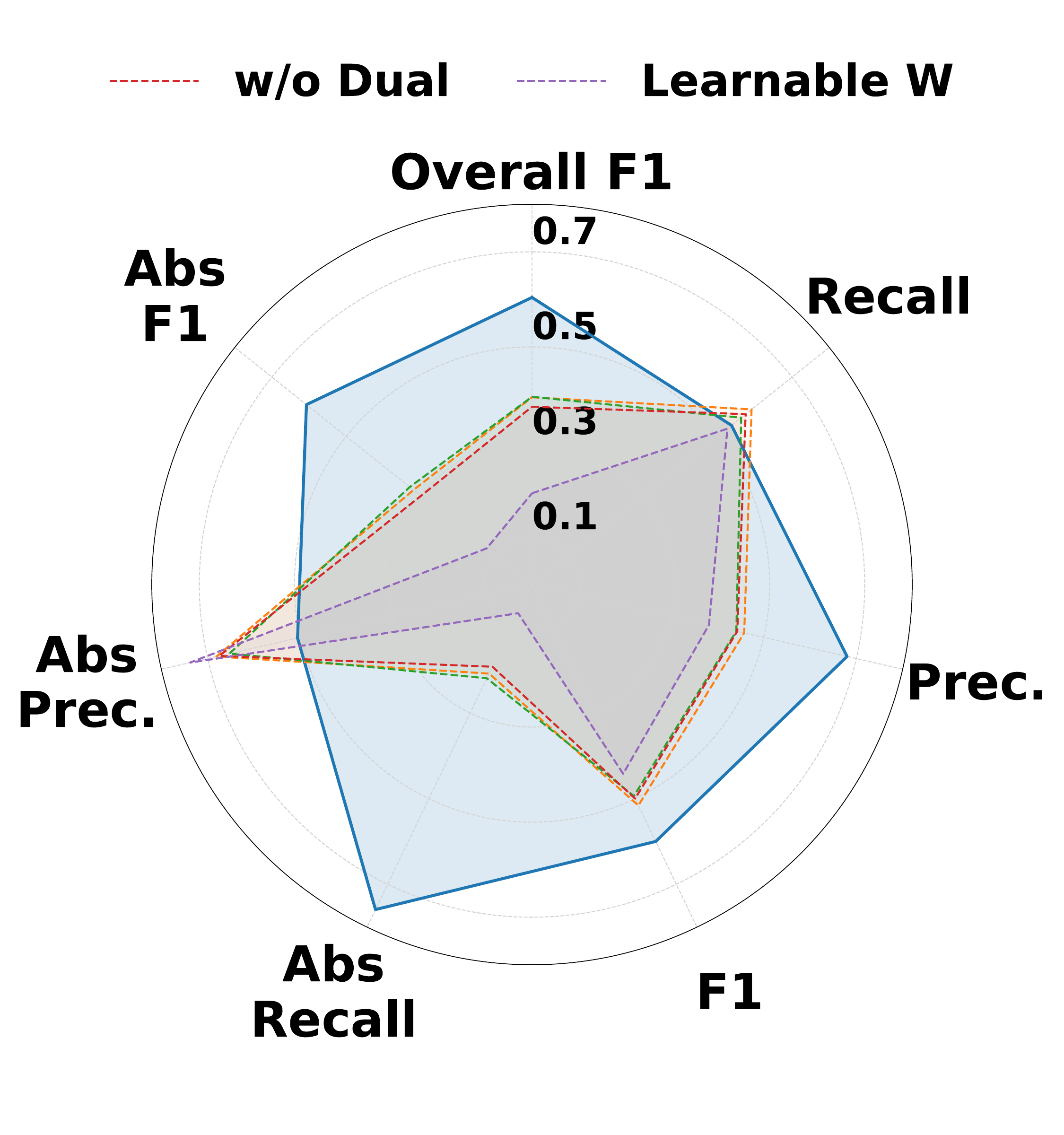}
        \caption{Llama3.1-8B}
    \end{subfigure}
    \cutcaptionup
    \caption{Performance comparison of ablation studies. The consistent degradation in Overall F1 scores observed when removing or altering individual components validates the distinct contribution of each module in the \model{} framework. }
    \label{fig:ablation}
\end{figure}

\subsection{Component Analysis}
\mypara{Ablation Study} We conduct an ablation study to evaluate the contribution of each component in our full model (\model{}). The following variations are compared: (1) w/o Dual removes the parametric memory head, relying solely on RAG context and thereby bypassing the internal-external conflict detection mechanism; (2) Learnable W replaces the evidence-based Dempster-Shafer fusion with a simple learnable scalar gate to weight the loss; (3) CE Only substitutes the evidential learning objective with standard cross-entropy loss, ignoring explicit uncertainty quantification; and (4) w/o KL omits the KL divergence regularization term from the evidential loss, utilizing only the data fit term.

Figure~\ref{fig:ablation} shows that the full model with all components performs the best across all metrics. Specifically, the removal of the dual-head architecture (w/o Dual) results in a critical drop in Abstain F1, confirming that explicitly modeling the conflict between parametric and retrieved knowledge is essential for mitigating hallucinations. Furthermore, the Learnable W variant exhibits the lowest Trade-Off scores, indicating that simple scalar gating is insufficient for integrating conflicting evidence compared to our evidential Dempster-Shafer fusion. Finally, the sub-optimal performance of the CE baseline and the ablation without KL regularization highlights the necessity of evidential uncertainty quantification. Specifically, KL regularization is essential for effectively distinguishing true unknowns from data ambiguity, thereby ensuring robust calibration.\looseness=-1

\mypara{Qualitative Analysis} We first investigate how well our evidential framework calibrates uncertainty compared to standard optimization methods. To visualize this, we compare the uncertainty distributions and Expected Calibration Error (ECE) between the Cross-Entropy (CE) baseline and our Evidential Deep Learning (EDL) approach. Figure~\ref{fig:uncertainty} illustrates the simplex of prediction probabilities, where CE (red) tends to cluster densely around high-confidence regions (Avg $P_1 \approx 0.87$), indicating a strong tendency toward overconfidence even when the model is likely wrong. In contrast, EDL (blue) exhibits a more dispersed distribution with a lower average confidence (Avg $P_1 \approx 0.77$), suggesting a more conservative and reliable estimation of belief. This improved calibration is quantitatively supported by Figure~\ref{fig:ece_error}, where EDL consistently achieves substantially lower ECE scores across all four knowledge quadrants (KG, KN, UG, UN). Notably, in the high-risk Unknown-Noisy (UN) and Known-Noisy (KN) scenarios, EDL significantly reduces calibration error, demonstrating its effectiveness in mitigating the overconfidence typically observed in hallucinating models.

Next, we examine the model's sensitivity to knowledge conflicts—specifically, determining whether the system can effectively distinguish between scenarios where its internal parametric memory aligns with external retrieved evidence versus when they contradict. To quantify this, we utilize the Jensen-Shannon (JS) divergence to measure the distributional discrepancy between the predictive distributions derived from the parametric and retrieval-augmented paths. Figure~\ref{fig:js_div} reveals that the CE baseline fails to capture this distinction, showing a negligible increase in divergence ($+0.005$) when a conflict occurs. Conversely, our method demonstrates a marked sensitivity to discordance, with a significant increase in JS divergence ($+0.039$) for conflicting samples compared to consistent ones. This suggests that our evidential framework effectively disentangles aleatoric uncertainty from epistemic belief, allowing the model to explicitly detect conflicts and thereby filter out contaminated knowledge more effectively than standard probabilistic approaches.\looseness=-1

\section{Conclusion}
This paper proposes \model{}, a principled framework designed to address the critical challenge of knowledge conflicts in RAG systems. We propose shifting the paradigm of confidence estimation from scalar probabilities to evidence distributions via Evidential Deep Learning and introduced a rigorous geometric quantification of "belief conflict" using Dempster-Shafer Theory. With these components, our method successfully disentangles epistemic uncertainty from aleatoric ambiguity and explicitly measures the discordance between its internal parametric knowledge and external retrieved evidence. As a result, our method produces well-calibrated confidence estimates and significantly outperforms existing baselines in reducing hallucination risks while maintaining high accuracy on answerable queries. We hope that these findings underscore the necessity of modeling evidence distributions, laying the groundwork for future research into trustworthy AI systems.

\clearpage

\bibliographystyle{ACM-Reference-Format}
\bibliography{main}

\newpage
\appendix
\section{Appendix}
\subsection{Training Details} We release the source code and dataset for \model{} at~\url{https://anonymous.4open.science/r/ERA-318B}.

\mypara{Dataset Details} To rigorously evaluate our framework, we first construct a dataset categorized by the interplay between the model’s internal knowledge state and the quality of external retrieval. We process standard QA datasets using the target LLM to establish its parametric knowledge boundary, classifying each query as Known or Unknown based on the model's ability to answer correctly without retrieval.  To determine membership in $\mathcal{K}_{\text{param}}$, we set the consistency threshold $\delta$ to 1.0 using $N=10$ sampled responses. 
Subsequently, for each query, we retrieve a pool of candidate contexts containing a mix of relevant ("golden") and irrelevant ("noisy") passages. Based on the presence of golden contexts in the top-3 retrieved results, we stratify the data into four distinct quadrants. Following prior work, we randomly sample 5,000 examples for training and 3,000 for evaluation from these stratified quadrants across three datasets, setting the IDK-ratio hyperparameter to 0.7 for the training distribution.

To further assess the generalizability of RAG models against noisy retrieval contexts, we constructed a synthetic dataset derived from the Wiki Event corpus. We employed GPT-4o to generate context-dependent questions and short answers directly from the original reference documents. To simulate realistic and challenging retrieval scenarios, each query is paired with a stratified pool of 10 contexts designed to represent varying levels of relevance. This pool comprises a single Golden Context containing the ground truth, four Hard Negatives generated by the LLM to exhibit high topical similarity and keyword overlap while explicitly excluding the correct answer, and five Easy Negatives randomly sampled from the global corpus to represent irrelevant noise. This hierarchical structure allows for a granular assessment of the model's ability to discern correct evidence amidst varying degrees of semantic interference and distraction.

\mypara{Implementation Details}  We employ Qwen3-8B-Base and Llama-3.1-8B as the primary backbone models, optimized via 4-bit Quantized LoRA (QLoRA) with nf4 quantization and bfloat16 precision to ensure computational efficiency. LoRA adapters are applied to all linear layers with a rank $r=8$, $\alpha=16$, and a dropout rate of $0.05$. We train the model for 1 epoch using the AdamW optimizer with a cosine learning rate scheduler, initialized at $5e-5$ with a warmup ratio of $0.1$. For the \model{} framework, we introduce two lightweight evidential heads that map hidden states to 4-dimensional Dirichlet parameters using a Softplus activation. We utilize the Digamma loss with KL divergence regularization, applying an annealing schedule over the first $2,000$ steps for stability. The balancing coefficients $\lambda_{\text{rag}}$ and $\lambda_{\text{param}}$ are set to $0.1$, while the conflict-aware weighting factor $\gamma$ for Dempster-Shafer combination is set to $1.0$. Additionally, a standard SFT auxiliary loss with a coefficient of $1.0$ is incorporated to maintain generation quality. All experiments are performed on 4 NVIDIA A100 GPUs using PyTorch. 

\subsection{Further Results on Honest Performance} 

Due to space limitations in the main manuscript, we present additional results here.

\begin{figure}[!t] 
        \centering \includegraphics[width=\linewidth]{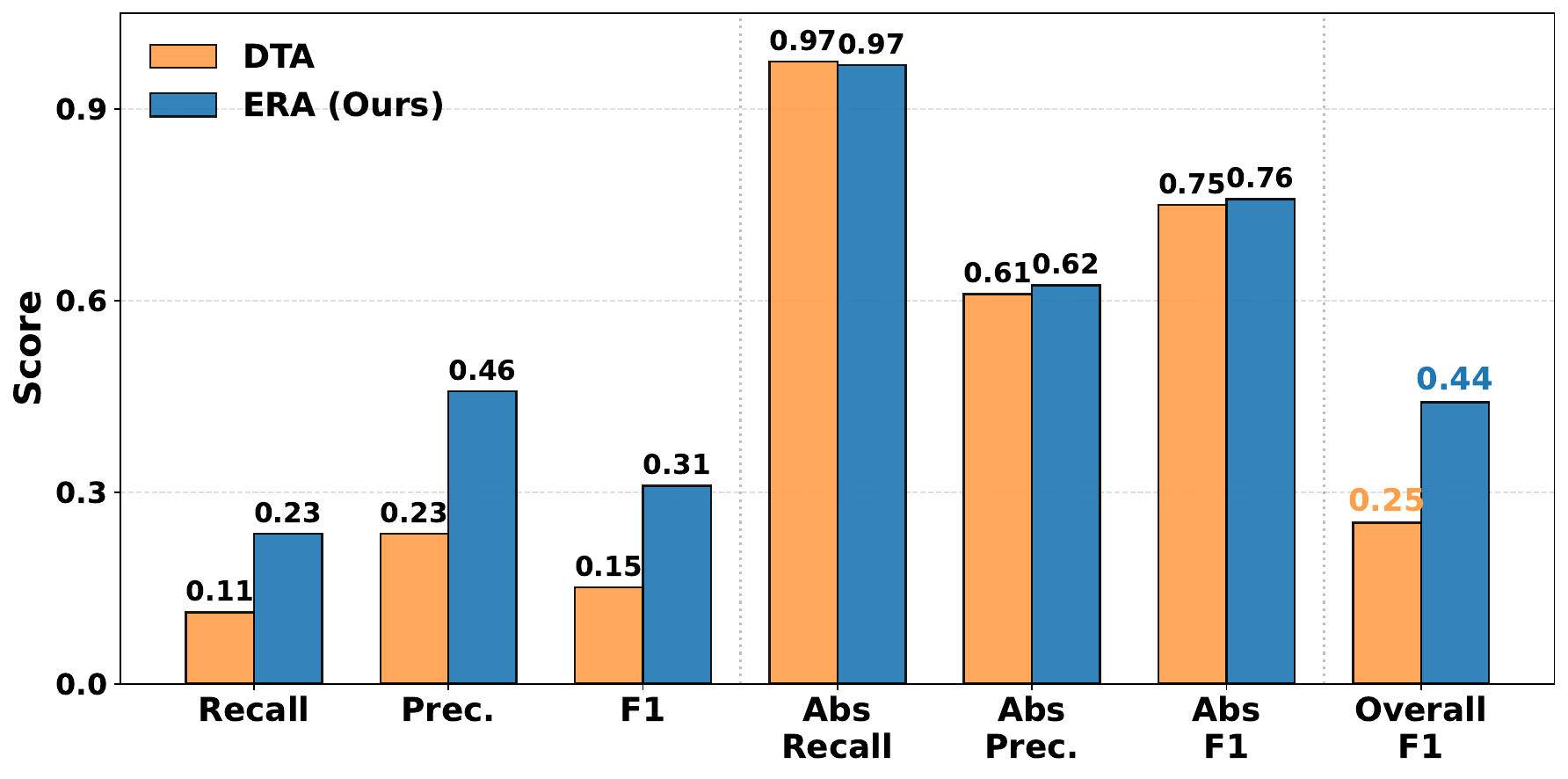}
        \caption{Generalizability evaluation on the Wiki Event dataset using Llama-3.1-8B.}
    \label{fig:wiki_result_llama}
\end{figure}

\mypara{Generalization Analysis} As illustrated in Figure~\ref{fig:wiki_result_llama}, our evaluation on Llama-3.1-8B highlights the superior generalization capability of \model{}. While baselines like DTA achieve high abstention recall, they demonstrate a tendency toward over-conservatism, indiscriminately rejecting answerable In-Distribution (ID) queries alongside OOD noise. In contrast, \model{} establishes a precise decision boundary that effectively distinguishes between ID and OOD inputs. By strictly refusing OOD inputs without compromising the ability to answer valid queries, \model{} achieves a marked improvement in overall F1 scores, demonstrating robust generalization across mixed distributions.

\begin{figure}[!t] 
        \centering \includegraphics[width=\linewidth]{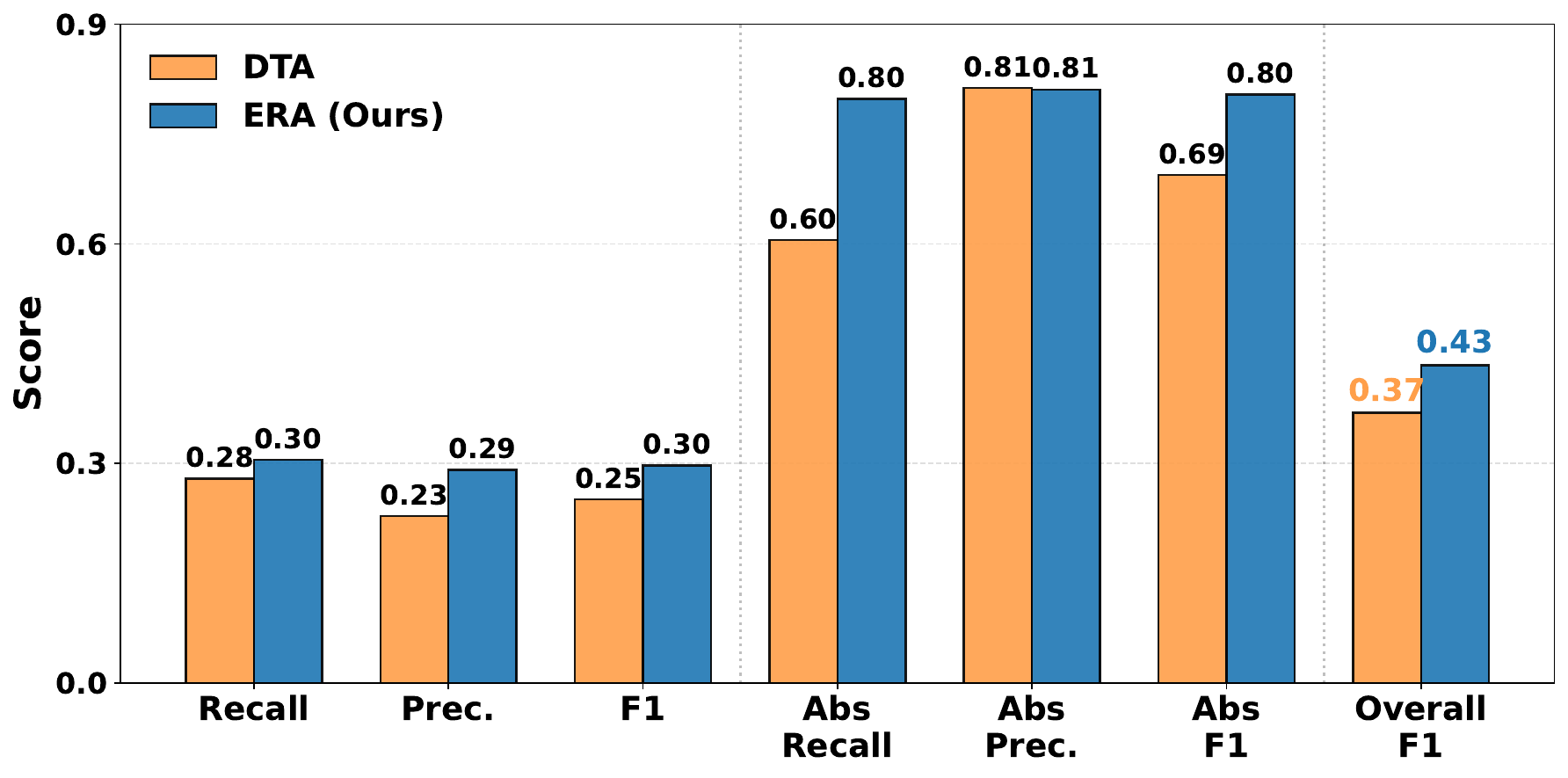}
        \caption{Generalizability evaluation on the Wiki Event dataset using ChatQA.}
    \label{fig:wiki_result_chatqa}
\end{figure}

\mypara{Add-on to RAFT} The performance advantage extends to models explicitly optimized for Retrieval Augmented Fine-Tuning (RAFT), such as ChatQA. As shown in Figure~\ref{fig:wiki_result_chatqa}, applying \model{} to these architectures yields substantial improvements in standard performance metrics compared to DTA. Building upon their pre-training for retrieval contexts, integrating \model{} further refines their ability to discriminate between relevant and irrelevant information.

\end{document}